\documentclass[reprint,nofootinbib,amsmath,amssymb,aps]{revtex4-2}

\usepackage[english]{babel}
\usepackage{booktabs}
\usepackage{graphicx}
\usepackage{hyperref}
\usepackage[margin=25mm]{geometry}

\hypersetup{colorlinks=true,linkcolor=blue,citecolor=blue,urlcolor=blue}

\begin{document}

\title{Catalog of electroweak scalar manifolds}

\author{Juan Carlos Criado}
\affiliation{Departamento de Física Teórica y del Cosmos,
  Universidad de Granada, Campus de Fuentenueva, E--18071 Granada, Spain}

\begin{abstract}
  The local structure of the Higgs sector around the vacuum does not uniquely determine its global properties.
  Most of the current experimental data provides only local information, which allows for a rich variety of global features, including several distinct topologies of the scalar manifold, and the existence of zero, one, or two fixed points of the symmetry transformations.
  Here, I provide, under general conditions, a complete classification of realizations of the electroweak symmetry with minimal field content---the three would-be Goldstone bosons and the Higgs---and outline some of their physical consequences.
\end{abstract}

\maketitle

\section{Introduction}

The Higgs Effective Field Theory (HEFT) is a general framework for the description of physics at energies around the electroweak scale~\cite{Feruglio:1992wf,Alonso:2012px,Buchalla:2013rka,Brivio:2017vri}.
It contains 4 scalar fields parametrizing a 4-dimensional target manifold $M$, on which the electroweak symmetry group acts smoothly.
The Coleman-Callan-Wess-Zumino (CCWZ) construction~\cite{Coleman:1969sm,Callan:1969sn} implies that the collection of vacua forms a 3-dimensional submanifold diffeomorphic to a 3-sphere.%
\footnote{Other 3-dimensional manifolds are allowed by the known infinitesimal transformation properties of the scalars, but they present problems in the description of essential features of the electroweak theory, such as the fermion masses~\cite{Gripaios:2016ubw}.}
The 3 Goldstone modes take values on this 3-sphere, while the Higgs field corresponds to the direction transverse to it.%
\footnote{The name ``Higgs field'' is used here to refer to the field that describes the spin-0 particle discovered at the Large Hadron Collider~\cite{ATLAS:2012yve,CMS:2012qbp}, and not necessarily to a specific component of the scalar fields involved in a Higgs mechanism}
The electroweak symmetry acts on the Goldstones, leaving the Higgs invariant.

The Standard Model (SM) is a particular case of this setup.
The target manifold $M$ is a vector space, on which the symmetries act linearly.
The vacuum submanifold is a 3-sphere around the origin, whose radius is the Higgs vacuum expectation value.
The Higgs field parametrizes the radial direction.
A notable feature of the linear representation is the presence of a fixed point under the action of the symmetry transformations, localized at the origin~\cite{Alonso:2015fsp,Alonso:2016oah}.
However, the existence of a fixed point is not unique to this case, and in non-linear realizations one may find 0, 1, or 2 fixed points, as shown in this paper.

Apart from the topological and group-theoretical structure of $M$, which is the focus of this work, the HEFT Lagrangian endows it with additional geometrical information.
This Lagrangian may contain singularities, such as poles and branching points, at certain locations of the scalar target space~\cite{Falkowski:2019tft, Alonso:2021rac}.
I will consider $M$ to be the result of removing all singular points from this space.
This is the adequate treatment for a (differential) topological study: the fields should not be able to reach the singular points, and the symmetries should act smoothly on the rest of the target space.

In this article, I study the possible topologies that $M$ might have, and the corresponding actions of the electroweak symmetry group on them.
The available local information around the vacuum is sufficiently restrictive to allow only for a few possibilities.

The rest of this paper is organized as follows.
In Section~\ref{sec:classification}, I review the relevant theory of group actions on manifolds, illustrate it with a toy example, and apply it to the classification of HEFT scalar manifolds.
In Section~\ref{sec:physics}, I briefly discuss some of the physical consequences of the choice of manifold.
The conclusions are presented in Section~\ref{sec:conclusions}.

\section{Classification}
\label{sec:classification}

\subsection{Definitions and general results}

Let $G$ be a compact connected Lie group.
An action of $G$ on a manifold $M$ is a smooth map $G \times M \to M$, denoted as $(g, \phi) \mapsto g \cdot \phi$, that preserves the group structure: $g_1 \cdot (g_2 \cdot \phi) = g_1g_2 \cdot \phi$ and $1 \cdot \phi = \phi$.
Such a map defines an equivalence relation in $M$ given by $\phi_1 \sim \phi_2$ if there is some $g \in G$ such that $\phi_1 = g \cdot \phi_2$.
The equivalence classes of this relation are the orbits of the action.
The set of all orbits is denoted as $M / G$, which has a topological space structure induced by the projection $M \to M/G$.

At each point $\phi \in M$, the isotropy group is the set of all group elements $g$ that leave $\phi$ invariant:
\begin{equation}
  G_\phi = \{g \in G: g \cdot \phi = \phi\}.
\end{equation}
The isotropy groups of any two points in the same orbit are conjugate to each other in $G$, and thus isomorphic.
Any orbit is diffeomorphic to the space $G/G_\phi$ where $\phi$ is any point in it.

It can be shown that there is always a minimal isotropy group $H$, contained in all other isotropies (up to conjugation in $G$)~\cite{zbMATH03106034,mostert1957compact}.
This is known as the principal isotropy.
The corresponding orbits, which are diffeomorphic to $G/H$, are also called principal.
They have the maximal dimension among all orbits.
The non-principal isotropy groups and orbits are called singular.

In the applications considered in this paper, the principal orbits have codimension one.
One then says that the action has cohomogeneity one.
If $M$ is connected, the orbit space $M/G$ must be one of 4 possibilities~\cite{mostert1957compact}:
\begin{equation}
  S^1, \quad \mathbb{R}, \quad [0, \infty), \quad \text{or} \quad [0, 1].
\end{equation}
%
If the orbit space is a circle $S^1$ or a line $\mathbb{R}$, all orbits must be principal.
For a ray $[0, \infty)$ or an interval $[0, 1]$, the orbits in the interior are principal, while the ones at the endpoints should be singular.

The manifold $M$ and the $G$ action on it are fully specified by the orbit space $M/G$, the principal isotropy group $H$, and the singular isotropies $K_0$ and $K_1$ at the endpoints, if they are present.
In fact, $M$ can be constructed as~\cite{mostert1957compact}:
\begin{equation}
  M \cong (M/G \; \times \; G/H) / \sim,
  \label{eq:manifold-construction}
\end{equation}
where, if $p_i$ is an endpoint of $M/G$ and $K_i$ is the corresponding isotropy, the equivalence relation is given by $(p_i, g_1 H) \sim (p_i, g_2 H)$ whenever there is some $k\in K_i$ such that $g_1 = k g_2$.

In the cases without endpoints, $M$ is a simple product of the orbit space (a circle or a line) with the principal orbit~\cite{parker19864}.
In the $[0, \infty)$ case, the space $M/G \times G/H$ looks like a semi-infinite tube, with section $G/H$.
The equivalence relation squashes the boundary $\{0\} \times G/H$ to $\{0\} \times G/K_0$.
Schematically:
\begin{center}
  \includegraphics[width=0.9\linewidth]{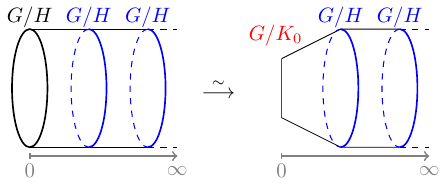}
\end{center}
where the blue circles represent the principal orbits, and the red segment represents the singular one.

To see what happens in the $[0, 1]$ case, consider the compactification of the $[0, \infty)$ manifolds by adding a point at infinity:
\begin{center}
  \includegraphics[width=\linewidth]{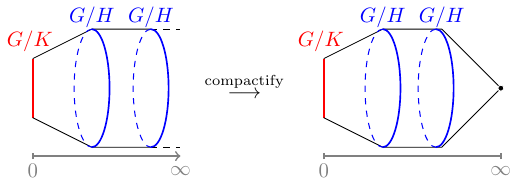}
\end{center}
All of the manifolds with $[0, 1]$ orbit space can be obtained by cutting two of the compactified $[0, \infty)$ ones $M_1$ and $M_2$ at some principal orbit, and gluing them along it.
To define the gluing, one needs a diffeomorphism between the boundaries $f: G/H \to G/H$, compatible with the group action: $f(g \cdot \phi) = g \cdot f(\phi)$.
Such an $f$ is said to be equivariant.
A common special case arises when the principal orbit is an $n$-sphere $S^n$.
The cutting and gluing operation is then the connected sum, denoted as $M_1\# M_2$.
In pictures:
\begin{center}
  \includegraphics[width=0.9\linewidth]{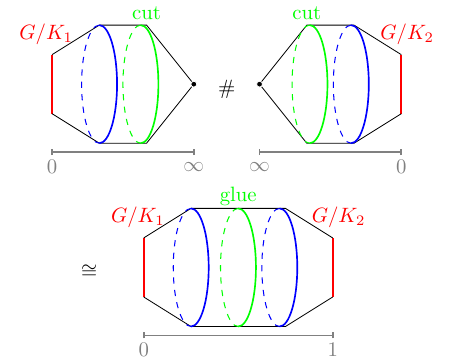}
\end{center}
In principle, there may be several options for the diffeomorphism $f$.
In many instances, all of them give the same manifold, even before imposing compatibility with the group action.
This will be the case in most of the applications I will consider, except one: the connected sum of the complex projective space $\mathbb{C}P^2$ with itself, for which there are two choices, as discussed below.

More information about the singular isotropy groups $K_i$ is provided by the following property:
$K_i/H$ must be an $n$-sphere $S^n$ for some $n$, including $S^0 \cong \mathbb{Z}_2$~\cite{mostert1957compact}.
When $H$ is trivial, this means that $K_i$ itself is an $n$-sphere.

\subsection{Toy example: $U(1)$}

\begin{figure*}
  \centering
  \includegraphics[width=0.9\textwidth]{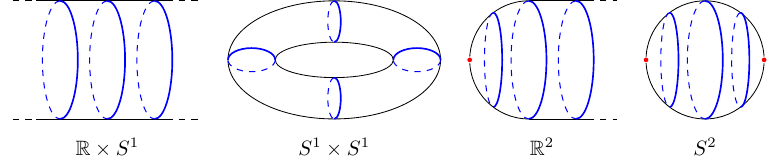} \\[10pt]
  \includegraphics[width=0.8\textwidth]{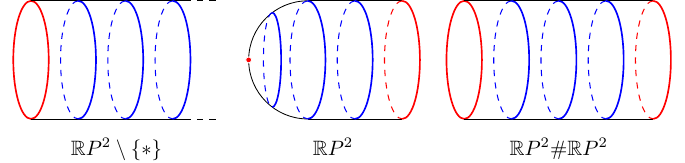}
  \caption{The seven possible actions of $U(1)$ on 2-dimensional surfaces. Samples of the circle ($S^1$) principal orbits are shown as thicker blue lines, while the fixed points are displayed as red dots. The red circles correspond to singular circle orbits in which opposite points should be identified. Some actions of the electroweak symmetry group on 4-dimensional surfaces can be obtained by replacing the principal orbits by 3-spheres and the singular ones by projective planes $\mathbb{R}P^3$, which is given by identifying opposite sides in a 3-sphere.}
  \label{fig:surfaces}
\end{figure*}

To introduce most of the features of the full problem in a simpler setup, I will consider first a toy version.
Suppose the full symmetry group is $U(1) \cong S^1$, spontaneously broken to the trivial one.
The vacuum manifold is then a circle $S^1$, parametrized by the single Goldstone.
With an additional Higgs field, the scalar target manifold $M$ is a 2-dimensional surface that looks like $\mathbb{R} \times S^1$ near the vacuum, with $U(1)$ acting freely on the $S^1$ factor.
So the action has cohomogeneity one, the principal isotropy is trivial, and the principal orbit is $S^1$.

Eq.~\eqref{eq:manifold-construction} (and the procedure described below it) can now be used to obtain all surfaces $M$ with such an action, which are given in Table~\ref{tab:toy}.
A schematic representation of the actions over all seven possible surfaces is shown in Fig.~\ref{fig:surfaces}.
Although this classification has been presented in the mathematical literature, I briefly review its derivation below, for completeness.

When the orbit space is $S^1$, the surface must be a torus $S^1 \times S^1$.
Similarly, when it is $\mathbb{R}$, the corresponding surface is a cylinder $\mathbb{R} \times S^1$.

For the $[0, \infty)$ orbit space, the surface is a semi-infinite cylinder $[0, \infty) \times S^1$, in which some identification has been performed in the boundary circle $\{0\} \times S^1$.
The possible isotropy groups at the boundary are either the finite discrete subgroup $\mathbb{Z}_2$, or the full group $U(1)$.
$\mathbb{Z}_2$ identifies opposite points of the circle, generating a Möbius strip.
When the isotropy is the full group, the boundary circle is identified to a single point, and the action becomes the linear action of $U(1)$ over the plane $\mathbb{R}^2$.

Finally, when the orbit space is an interval $[0, 1]$, there are 3 choices of isotropy groups at the endpoints: $(K_0, K_1) = (\mathbb{Z}_2, \mathbb{Z}_2)$, $(\mathbb{Z}_2, U(1))$, or $(U(1), U(1))$.
The corresponding manifolds are obtained as connected sums of the compactifications of the $[0, \infty)$ surfaces, which are a proyective plane $\mathbb{R}P^2$ (for the Möbius strip) and a sphere $S^2$ (for the plane $\mathbb{R}^2$).
Therefore, the 3 choices of isotropy groups lead to:
\begin{itemize}
    \item $M = \mathbb{R}P^2 \# \mathbb{R}P^2$, a Klein bottle.
    \item $M = \mathbb{R}P^2 \# S^2 \cong \mathbb{R}P^2$,  the projective plane.
    \item $M = S^2 \# S^2 \cong S^2$, the sphere.
\end{itemize}
This completes the classification.

This toy example captures most of the features that will appear in the main case.
There is a variety of surfaces with different homotopy properties.
They can be both compact or non-compact, and orientable or non-orientable.
The action can have any number of fixed points from 0 to 2, and there are both linear (on $\mathbb{R}^2$) and non-linear (on $\mathbb{R}P^2$) actions with 1 fixed point.

\begin{table}
  \centering
  \begin{tabular}{cccclc}
    \toprule
    Orbit space & $K_0$ & $K_1$ & $M$ & Name & $n_{\rm fix}$ \\
    \midrule
    $S^1$         & --             & -- & $S^1 \times S^1$ & torus              & 0 \\
    $\mathbb{R}$  & --             & -- & $\mathbb{R} \times S^1$ & cylinder       & 0 \\
    $[0, \infty)$ & $\mathbb{Z}_2$ & -- & $\mathbb{R}P^2 \setminus \{*\}$ & Möbius strip                        & 0\\
    $[0, \infty)$ & $S^1$          & -- & $\mathbb{R}^2$ & plane                & 1 \\
    $[0, 1]$      & $\mathbb{Z}_2$ & $\mathbb{Z}_2$ & $\mathbb{R}P^2 \# \mathbb{R}P^2$
                                      & Klein bottle & 0 \\
    $[0, 1]$      & $\mathbb{Z}_2$ & $U(1)$ & $\mathbb{R}P^2$ & proj. plane      & 1 \\
    $[0, 1]$      & $U(1)$ & $U(1)$  & $S^2$ & sphere                             & 2 \\
    \bottomrule
  \end{tabular}
  \caption{Classification of the actions of $U(1)$ on a connected 2-dimensional manifold with principal orbit $S^1$. $n_{\rm fix}$ is the number of fixed points.}
  \label{tab:toy}
\end{table}

\subsection{Electroweak scalar manifolds}
\label{sec:electroweak}

The electroweak symmetry group is $G = SU(2)_L \times U(1)_Y$.
The target space for the scalars is a 4-dimensional manifold $M$ on which $G$ acts with principal isotropy $H = U(1)_{\rm em}$, the electromagnetic (diagonal) subgroup.
This generates principal orbits that are 3-spheres $G/H \cong S^3$.
The results of Ref.~\cite{hoelscher2010classification} imply that every such action gives rise to an action on the same manifold $M$ of the group $G = SU(2)$ with trivial principal isotropy $H = \{1\}$, and the same principal orbit $G/H \cong S^3$.
This is called a reduced action.
Under mild conditions, a reduced action does also generate an action of the original group.
These conditions will be met in the classification provided in this section.
For these reasons, I will consider reduced actions in what follows, without loss of generality.
In Appendix~\ref{app:actions}, I describe how to recover the action of $SU(2)_L \times U(1)_Y$.

As before, I will proceed by considering each possible orbit space separately.
Similarly to the $U(1)$ case, when the orbit space is a circle or a line, the manifold $M$ is either $S^1 \times S^3$ or $\mathbb{R} \times S^3$.

The singular isotropies are the subgroups of $SU(2)$ diffeomorphic to an $n$-sphere: $\mathbb{Z}_2$, $U(1)$ or the full $SU(2)$.
The corresponding manifolds with orbit space $[0, \infty)$ are a punctured real projective space $\mathbb{R}P^4 \setminus \{*\}$, a punctured complex projective space $\mathbb{C}P^2 \setminus \{*\}$ and the vector space $\mathbb{R}^4$.
Their compactifications are $\mathbb{R}P^4$, $\mathbb{C}P^2$ and $S^4$.
The manifolds with orbit space $[0, 1]$ are their connected sums.
There are 6 combinations:
\begin{align}
  &\mathbb{R}P^4 \# \mathbb{R}P^4, \qquad
  &&\mathbb{R}P^4 \# \mathbb{C}P^2, \nonumber \\
  &\mathbb{R}P^4 \# S^4 \cong \mathbb{R}P^4, \qquad
  &&\mathbb{C}P^2 \# \overline{\mathbb{C}P^2}, \\
  &\mathbb{C}P^2 \# S^4 \cong \mathbb{C}P^2, \qquad
  &&S^4 \# S^4 \cong S^4. \nonumber
\end{align}
The complete list, with the corresponding isotropy groups and the counting of fixed points is given in Table~\ref{tab:classification}.
The subset of compact manifolds has been computed before in Ref.~\cite{parker19864}.
In Appendix~\ref{app:actions}, I provide additional details about the action of the electroweak group on the manifolds in the classification.

As noted above, the gluing procedure in a connected sum requires the specification of a diffeomorphism $f: S^3 \to S^3$.
In all cases except for the sum of $\mathbb{C}P^2$ with itself, all diffeomorphisms give rise to the same manifold.
There are two ways to perform connected sum of $\mathbb{C}P^2$ with itself, corresponding to the two homotopy classes of mappings $S^3 \to S^3$ with winding number +1 and -1.
The notation $\mathbb{C}P^2 \# \overline{\mathbb{C}P^2}$ refers to the case in which one picks an orientation in $\mathbb{C}P^2$ and chooses a diffeomorphism $f$ that preserves the orientation it induces on the $S^3$ cuts.
The other possibility is denoted $\mathbb{C}P^2 \# \mathbb{C}P^2$, and it is not compatible with the group action~\cite{parker19864}.

\begin{table}
  \centering
  \begin{tabular}{ccccc}
    \toprule
    Orbit space & $K_0$ & $K_1$ & $M$  & $n_{\rm fix}$ \\
    \midrule
    $S^1$         & --             & -- & $S^1 \times S^3$& 0 \\
    $\mathbb{R}$  & --             & -- & $\mathbb{R} \times S^3$ & 0 \\
    $[0, \infty)$ & $\mathbb{Z}_2$ & -- & $\mathbb{R}P^4 \setminus \{*\}$ & 0\\
    $[0, \infty)$ & $U(1)$          & -- & $\mathbb{C}P^2 \setminus \{*\}$ & 0 \\
    $[0, \infty)$ & $SU(2)$          & -- & $\mathbb{R}^4$ & 1 \\
    $[0, 1]$      & $\mathbb{Z}_2$ & $\mathbb{Z}_2$ & $\mathbb{R}P^4 \# \mathbb{R}P^4$ & 0 \\
    $[0, 1]$      & $\mathbb{Z}_2$ & $U(1)$ & $\mathbb{R}P^2 \# \mathbb{C}P^2$     & 0 \\
    $[0, 1]$      & $\mathbb{Z}_2$ & $SU(2)$ & $\mathbb{R}P^4$     & 1 \\
    $[0, 1]$      & $U(1)$ & $U(1)$  & $\mathbb{C}P^2 \# \overline{\mathbb{C}P^2}$            & 0 \\
    $[0, 1]$      & $U(1)$ & $SU(2)$  & $\mathbb{C}P^2$                            & 1 \\
    $[0, 1]$      & $SU(2)$ & $SU(2)$  & $S^4$                            & 2 \\
    \bottomrule
  \end{tabular}
  \caption{Classification of the actions of $SU(2)$ on a 4-dimensional connected manifold with principal orbit $S^3$. The column labeled $M$ lists all the possible target manifolds for the scalar fields in the HEFT.}
  \label{tab:classification}
\end{table}

It is worth compiling the list of possible singular orbits that may appear.
For singular isotropy $K_i = \mathbb{Z}_2$, they are $SU(2) / \mathbb{Z}_2 \cong \mathbb{R}P^3$.
For $K_i = U(1)$, they are $SU(2) / U(1) \cong S^2$.
For $K_i = SU(2)$, they are fixed points $SU(2) / SU(2) \cong \{*\}$.

Near a fixed point, the action always looks like a linear one.
That is, there is an equivariant diffeomorphism $f$ from a neighborhood of the fixed point in $M$ to a neighborhood of the origin in the vector space $\mathbb{R}^4$, with $SU(2)$ acting on $\mathbb{R}^4 \cong \mathbb{C}^2$ as the fundamental representation.
In general, this cannot be extended the complete manifold $M$, unless $M = \mathbb{R}^4$.

Finally, I emphasize here that Table~\ref{tab:classification} provides the list of all possible target manifolds for the scalar fields in the HEFT, assuming that the electroweak symmetry group acts smoothly on them, and that they look like a product $\mathbb{R} \times S^3$ near the vacuum.%
\footnote{Table~\ref{tab:classification} assumes that $M$ is connected. Dropping this requirement allows for arbitrary disjoint unions of the listed manifolds.\label{fn:connected}}
That is, it is assumed that in some region around the vacuum, the target manifold is parametrized by a Higgs field $h \in \mathbb{R}$, and the Goldstone matrix $U \in SU(2) \cong S^3$.
Each of the manifolds in the list gives rise to a different physical theory.
In principle, all of these theories are compatible with the current knowledge of the electroweak interactions, and so they must be included in a fully general Effective Field Theory (EFT).

\subsection{Physical conditions}

Beyond topology and group theory, one might wonder if there are any other obstructions to the construction of a physical theory in any of the manifolds in the classification.
From a bottom-up EFT point of view, it is sufficient that a Lagrangian reproducing the measured couplings exists.
Minimally, for the scalar sector, such a Lagrangian must contain a potential and a 2-derivative term.
The potential is a smooth function $V: M \to \mathbb{R}$, and the 2-derivative term can be constructed from a Riemannian metric $g$ on $M$~\cite{Alonso:2015fsp}.
In any coordinate patch with coordinates $\phi = (\phi_0, \phi_1, \phi_2, \phi_3)$, the Lagrangian takes the form
\begin{equation}
  \mathcal{L} = V(\phi) + g_{ij}(\phi) \partial_\mu \phi_i \partial^\mu \phi_j.
  \label{eq:generic-lagrangian}
\end{equation}

Both the potential $V$ and the metric $g$ must be invariant under the action of the electroweak symmetry group.
It is well-known that such objects exist in cohomogeneity one manifolds~\cite{galaz2018cohomogeneity}.
Away from singular orbits, one can choose coordinates $(h, U)$ as above.
Then, both objects depend only on $h$:
\begin{align}
  V(\phi) &= V(h), \qquad g(\phi) = dh^2 + g_{ij}(h) dL^i dL^j,
  \label{eq:V-and-g}
\end{align}
where $g_{ij}(h)$ is a symmetric positive-definite matrix, and
\begin{equation}
     dL^i = \langle \sigma^i U^\dagger d U\rangle
    \label{eq:dL-def}
\end{equation}
with $\sigma^i$ being the Pauli matrices, and angle brackets $\langle\cdot\rangle$ denoting the trace. As shown in Appendix~\ref{app:lagrangian},
the dependence on $h$ is arbitrary near the vacuum,%
\footnote{In fact, this is true near any principal orbit. At singular orbits (and thus away from the vacuum), there are smoothness conditions, also discussed in Appendix~\ref{app:lagrangian}.}
and it can thus be chosen so that it satisfies current experimental constraints.%
\footnote{See Ref.~\cite{Manton:2024eli} for a detailed study of the case $M = \mathbb{C}P^2$ with a concrete Lagrangian.}
Therefore, all manifolds in the list give rise to viable EFTs in a bottom-up approach.

It is harder to assess if every possibility is consistent with some ultraviolet (UV) completion.
For $M=\mathbb{C}P^2$, Ref.~\cite{Manton:2024eli} provides a discussion in this direction, but it is currently not known if they exist.
One approach is to consider a theory with an approximate global symmetry $\mathcal{G}$ spontaneously broken to $\mathcal{H}$, and interpret the associated pseudo-Goldstone bosons as the 4 scalar degrees of freedom in the electroweak theory.
Then, the electroweak scalar manifold is given by $M = \mathcal{G}/\mathcal{H}$.
This is possible for several of the manifolds in the list:
\stepcounter{footnote}
\begin{align}
  S^1 \times S^3 &\cong  (U(1) \times SU(2)) / \{1\}, \\
  \mathbb{R}P^4 &\cong  SO(5) / O(4), \\
  \mathbb{C}P^2 &\cong  SU(3) / U(2), \\
  S^4 &\cong  SO(5)/SO(4),\footnotemark[\value{footnote}]
\end{align}
\footnotetext[\value{footnote}]{This pattern appears in minimal composite Higgs models~\cite{Agashe:2004rs,Contino:2006qr}.}%
It is also clear that both $\mathbb{R} \times S^3$ and $\mathbb{R}^4$ do not present any topological obstructions to their UV completion.
The former has an $S^3$ factor that can be generated by spontaneous symmetry breaking as $SU(2)/\{1\}$ and a completely independent Higgs living in $\mathbb{R}$.
The latter is a linear realization.

Assuming the UV theory is weakly coupled and UV complete, it must implement a Higgs mechanism, in which the scalar fields are in a linear representation of the electroweak symmetry.
The manifold $M$ should then be embedded in the associated vector space, and the action on it should be the one inherited from the corresponding linear action.
Such an embedding is always possible~\cite{mostow1957equivariant}.
An immediate question is then if $M$ can be generated as a set of degenerate minima of some renormalizable UV potential $U$.
If such a $U$ exists, the full UV potential may be taken to be $V_{\rm UV} = U+ W$, where $W$ contains some relatively small corrections.
With an adequate hierarchy of scales between $U$ and $W$, one can integrate out the the degrees of freedom that go out of $M$, and obtain an EFT defined in $M$ with potential $V(\phi)$ given by the restriction of $W$ to $M$.
Some examples of $U$ are given in Table~\ref{tab:UV-examples}.
Additional details of how the extra degrees of freedom in the linear representation are integrated out and generate the EFT Lagrangian depend on the choice of $W$ and further details of the UV Lagrangian.
I will not study this in general here, but I provide an example for $M = S^4$ in Appendix~\ref{app:example}.

\begin{table}
    \begin{tabular}{ccc}
        \toprule
        Manifold & Fields & $U$ potential \\
        \midrule
        $\mathbb{R} \times S^3$ & $\begin{array}{c} h \in \mathbb{R} \\ \Phi \in \mathbb{R}^4\end{array}$ & $\omega (|\Phi|^2 - u^2)^2$
        \\[10pt]
        $S^1 \times S^3$ & $\begin{array}{c} \Phi_1 \in \mathbb{R}^2 \\ \Phi_2 \in \mathbb{R}^4\end{array}$ & $\begin{array}{c} \omega_1 (|\Phi_1|^2 - u_1^2)^2 \\ + \, \omega_2 (|\Phi_2|^2 - u_2^2)^2\end{array}$
        \\[10pt]
        $\mathbb{R}P^4$ & $\Phi \in \operatorname{Sym}(5)$ & $\begin{array}{c} \omega_1 u^2 \langle\Phi\rangle(\langle \Phi\rangle - 2 u) \\ + \, \omega_2 (\langle\Phi^2\rangle^2 - \langle \Phi^4\rangle)\end{array}$
        \\[10pt]
        $\mathbb{C}P^2$ & $\Phi \in \operatorname{Herm}(3)$ & $\begin{array}{c} \omega_1 u^2 \langle\Phi\rangle(\langle \Phi\rangle - 2u) \\ + \, \omega_2 (\langle\Phi^2\rangle^2 - \langle \Phi^4\rangle)\end{array}$
        \\[10pt]
        $S^4$ & $\Phi \in \mathbb{R}^5$ & $\omega (|\Phi|^2 - u^2)^2$ \\[5pt]
        \bottomrule
    \end{tabular}
    \caption{\label{tab:UV-examples} Examples of the UV potential $U$ having a set of degenerate minima diffemorphic to one of the manifolds in the classification. $\operatorname{Sym}(n)$ and $\operatorname{Herm}(n)$ denote the linear spaces of symmetric and Hermitian $n\times n$ matrices, respectively. $\omega_{(i)}$ and $u_{(i)}$ are free positive parameters.}
\end{table}

\section{Physical consequences}
\label{sec:physics}

\subsection{Local effects}

Although most of the physical effects of the choice of scalar manifold are related to its global topology, there are local effects associated with the infinitesimal action of the symmetries around singular orbits.
The physics of fixed points are well known: at them, the electroweak symmetry is restored, and so its 4 gauge bosons become massless.

The classification derived above shows that there is another possibility: if $SU(2) / U(1) \cong S^2$ orbits are present, there are points in $M$ at which the electroweak symmetry is only partially restored.
These points are not only invariant under the action of the generator of electromagnetic group $T_3 + Y$, but instead they remain invariant under the action of both $T_3$ and $Y$ separately.%
\footnote{This is because the singular isotropy is the maximal torus of $SU(2)_L \times U(1)_Y$, see Appendix~\ref{app:actions} for details.}
Then, both the $Z$ boson and the photon are massless, while the $W^\pm$ stay massive.

Near a principal orbit, the choice of $M$ does not restrict the Lagrangian in any way (see Appendix~\ref{app:lagrangian}).
In particular, one may set the Lagrangian to be the Standard Model one in a neighborhood of the vacuum.
Thus, the global topology does not have any consequences on the physical quantities determined by it, such as particle masses and couplings.
In Ref.~\cite{Manton:2024eli}, the maximally symmetric metric is chosen for $M = \mathbb{C}P^2$, allowing to derive predictions for these quantities.
A similar procedure might be applicable to other $M$.
However, if one adopts a bottom-up point of view, the form of the metric should be kept general, with the only restrictions to be applied being the ones derived from the available experimental data.
Then, the use of local information (for example, from collider experiments involving SM particles) does not allow to determine $M$.

\subsection{Global effects}
\label{sec:global}

\begin{table}
  \centering
  \begin{tabular}{ccccc}
    \toprule
    $M$ & $\pi_1(M)$ & $\pi_2(M)$ & $\pi_3(M)$ & $\pi_4(M)$ \\
    \midrule
    $S^1 \times S^3$ & $\mathbb{Z}$ & 0 & $\mathbb{Z}$ & $\mathbb{Z}_2$ \\
    $\mathbb{R} \times S^3$ & 0 & 0 & $\mathbb{Z}$ & $\mathbb{Z}_2$ \\
    $\mathbb{R}P^4 \setminus \{*\}$ & $\mathbb{Z}_2$ & 0 & $\mathbb{Z}$ & $\mathbb{Z}_2$ \\
    $\mathbb{C}P^2 \setminus \{*\}$ & 0 & $\mathbb{Z}$ & $\mathbb{Z}$ & $\mathbb{Z}_2$ \\
    $\mathbb{R}^4$ & 0 & 0 & 0 & 0 \\
    $\mathbb{R}P^4 \# \mathbb{R}P^4$ & $D_\infty$ & -- & -- & -- \\
    $\mathbb{R}P^4 \# \mathbb{C}P^2$ & $\mathbb{Z}_2$ & -- & -- & -- \\
    $\mathbb{R}P^4$ & $\mathbb{Z}_2$ & 0 & 0 & $\mathbb{Z}$ \\
    $\mathbb{C}P^2 \# \overline{\mathbb{C}P^2}$ & 0 & -- & -- & -- \\
    $\mathbb{C}P^2$ & 0 & $\mathbb{Z}$ & 0 & 0 \\
    $S^4$ & 0 & 0 & 0 & $\mathbb{Z}$ \\
    \bottomrule
  \end{tabular}
  \caption{Homotopy groups of the electroweak scalar manifolds.}
  \label{tab:homotopy}
\end{table}

The consequences of the global topology of $M$ are related to the existence of various topologically protected field configurations.
They can be understood through the homotopy groups $\pi_n(M)$, whose elements are the homotopy classes of mappings of $n$-spheres $S^n$ over $M$.
Two mappings are in the same class whenever they can be smoothly deformed into each other.
The first 4 homotopy groups for the manifolds in the classification of Table~\ref{tab:classification} are given in Table~\ref{tab:homotopy}.
I provide details on their computation in Appendix~\ref{app:homotopy}.
The missing elements in the table could not be calculated using the elementary techniques described in the Appendix.

On a superficial level, the scalar fields are mappings $\phi: \mathbb{R}^4 \to M$ from spacetime $\mathbb{R}^4$ to the target manifold $M$.
Such mappings are always homotopically trivial, because $\mathbb{R}^4$ is contractible.
Topological defects emerge when the mapping becomes singular in some lower-dimensional submanifold $T$ of $\mathbb{R}^4$.
One may view $T$ as the set of points at which the EFT fails, and thus a UV completion is needed to understand the microscopic details near them.
However, the EFT describes physics sufficiently far away from the singularities.
For this purpose, the fields should be treated as smooth mappings $\mathbb{R}^4 \setminus T \to M$.
A string has a two-dimensional worldsheet diffeomorphic to $\mathbb{R}^2$.
The corresponding domain is homotopically equivalent to a circle $\mathbb{R}^4 \setminus \mathbb{R}^2 \sim S^1$.
Thus, strings exist in a theory if $\pi_1(M)$ is non-trivial.
Similarly, monopoles have a one-dimensional worldline, and therefore they have an associated $\pi_2(M)$ class.
Instantons correspond to spacetime point defects, and so the relevant homotopy group for them is $\pi_3(M)$.

Physically, one regards the element of $\pi_n(M)$ associated with a topological defect as a topological charge $Q$ that takes discrete values.
The group structure corresponds to the operation of joining two defects initially far away.
The initial state is described by the class of each defect and is thus an element $(Q_1, Q_2)$ of $\pi_n(M) \times \pi_n(M)$.
The final state has the singular submanifolds $T_1$ and $T_2$ corresponding to both defects coinciding, and so it is an element $Q_1 + Q_2$ of $\pi_n(M)$.

The characterization of topological defects in terms of homotopy groups is typically done for the vacuum submanifold $V$ instead of the full target space $M$~\cite{Kibble:1976sj, Vilenkin:1984ib}.
This is just a particular case with $V = M$ of the previous discussion, which arises when the only degrees of freedom left in the EFT at hand are the Goldstones.
When non-vacuum points are included in $M$, the defects may have infinite action in infinite spacetime volume, or infinite energy in an infinite spatial slice.
In these cases, the field configuration should be viewed as a description of some bounded spacetime region around the defect.
This is also common in the usual analysis in terms of the vacuum manifold, because of non-vanishing gradients of the fields at $x \to \infty$.

\begin{figure*}
  \centering
  \includegraphics[width=\textwidth]{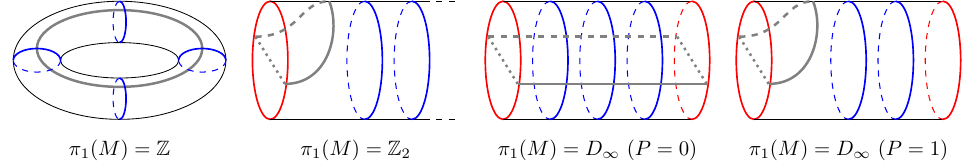}
  \caption{Schematic representation of closed paths (in gray) with topological charge $Q = 1$ in manifolds with different fundamental groups. Blue and red circles represent $S^3$ and $S^3/\mathbb{Z}_2 \cong \mathbb{R}P^3$ orbits, respectively.}
  \label{fig:paths}
\end{figure*}

The classification of strings that may emerge in the HEFT can be divided in 4 cases, corresponding to the 4 possible fundamental groups $\pi_1(M)$:
\begin{itemize}
    \item $\pi_1(M) = 0$: there are no topologically stable strings.
    \item $\pi_1(M) = \mathbb{Z}$: strings have an integer topological charge. This can only happen if $M = S^1 \times S^3$. Then, topologically-protected strings have the Higgs field, which lives on a circle, wrapping around them one or more times.
    \item $\pi_1(M) = \mathbb{Z}_2$: the topological charge is $Q \in \mathbb{Z}_2 \cong \{0, 1\}$. That is, there is only one type of string, with $Q = 1$. A string and an anti-string (with charge $Q = -1 \mod 2 = 1$) are the same object. This happens in spaces with a single $\mathbb{R}P^3$ orbit. Following the Higgs field through a circle around the string, it must touch this orbit an odd number of times.
  \item $\pi_1(M) = D_\infty$, the infinite dihedral group (see Appendix~\ref{app:homotopy} for its definition). This is the only example here of a non-Abelian fundamental group, which gives rise to strings that cannot pass through each other~\cite{Vilenkin:1984ib}. Strings have two associated charges: $Q \in \mathbb{Z}$ and $P \in \mathbb{Z}_2$. The one with $(Q, P) = (0, 1)$ touches only one of the $\mathbb{R}P^3$ orbits (an odd number of times), $(1, 1)$ only goes through the other one, and $(1, 0)$ connects them.
\end{itemize}
To illustrate the non-trivial paths along $M$ in each case, I display in Fig.~\ref{fig:paths} the examples of the paths with $Q = 1$.

The only non-trivial $\pi_2(M)$ that appears in the classification is $\mathbb{Z}$.
When it does, monopoles with integer topological $Q$ charge are present in the HEFT.
The image of a 2-sphere around the monopole maps surjectively to the singular $S^2$ orbit.

Instantons appear when $\pi_3(M) = \mathbb{Z}$.
These configurations $\phi : \mathbb{R}^4 \setminus \mathbb{R} \sim S^3 \to M$ can always be smoothly deformed so that $\phi$ takes values in the vacuum submanifold $V = S^3$.
In that case, the topogical charge becomes the winding number of the resulting $S^3 \to S^3$ map.%
\footnote{The $Q=1$ case corresponds with the map induced by the sphaleron at spacetime infinity~\cite{Manton:1983nd,Klinkhamer:1984di}.}
This is in contrast with strings and monopoles, which require the field values to leave the vacuum submanifold.

In addition to strings, monopoles, and instantons, there might be domain walls whenever $M$ is disconnected (see footnote~\ref{fn:connected}).
Throughout this paper, I assume a minimal HEFT vacuum manifold $V \cong S^3$.
If the Higgs potential had multiple degenerate minima, there would be several disconnected components of $V$, allowing for domain walls~\cite{Alonso:2023jsi}, as homotopically non-trivial mappings $\mathbb{R}^4 \setminus \mathbb{R}^3 \to V$.

A different type of topologically protected configuration appears by imposing that the fields are well defined everywhere in spacetime and have finite action.
This leads to boundary conditions that require them to go to a single point in $\phi_\infty \in V \cong S^3$ at spacetime infinity.
Then, spacetime can be compactified to $S^4$ by adding the point at infinity, while introducing the condition that there is some point $x_\infty \in S^4$ at which $\phi(x_\infty) = \phi_\infty$.
Non-trivial elements of $\pi_4(M)$ with representatives satisfying this are similar to instantons, but with a completely different origin as those defined above.
To differentiate them, I call those associated with $\pi_3$ instantons, and those related to $\pi_4$ spacetime textures.
These configurations are also associated with non-trivial loops in the space of spatial configurations of the fields~\cite{Finkelstein:1966ft}.

Spacetime textures arise in the HEFT in several forms.
For $M = S^4$ and $M = \mathbb{R}P^4$, the topological charge is an integer and the field configuration covers the complete manifold, so they probe all possible field values.
The remaining manifolds with non-trivial homotopy group have a $\mathbb{Z}_2$ charge, and the corresponding non-trivial spacetime textures admit a deformation in which their image is the vacuum $S^3$.
After this deformation, the field configuration must be a representative of the only non-trivial homotopy class of maps $S^4 \to S^3$.

For spatial configurations with finite energy, a similar reasoning leads to a domain for the fields that compactifies to $S^3$.
In this case, there are field configurations, usually known as textures, associated with non-trivial $\pi_3(M)$ classes.
When they are energetically stable, textures are called skyrmions~\cite{Skyrme:1961vq,Adkins:1983ya}.
They have been studied in the HEFT and some of its limits in Refs.~\cite{Ambjorn:1984bb,Ellis:2012cs,Kitano:2016ooc,Kitano:2017zqw,Criado:2020zwu,Hamada:2021oqm,Criado:2021tec,Bolognesi:2023gif}.
Just as instantons, when they exist, textures in the HEFT have integer topological charge and can always be deformed to take values in the vacuum $S^3$ only.

Additional considerations about the stability and general behaviour of the field configurations discussed in this section depend on the choice of Lagrangian and not only on the topological properties of $M$ and are therefore beyond the scope of this work.
Further structure appears when gauge fields are included in the discussion.
Topological defects can appear as combined configurations of scalar and vector bosons.
For simplicity, I have assumed here that the gauge fields are in a topologically trivial configuration.

\section{Conclusion}
\label{sec:conclusions}

The local structure around the vacuum of the HEFT scalar manifold, together with the condition that the electroweak symmetry group acts smoothly on it, are sufficient to determine a set of 11 possible manifolds and corresponding group actions.
Orbits diffeomorphic to a 3-sphere densely fill the manifold.
There can be 0, 1, or 2 isolated orbits of other types: fixed points, 2-spheres, or 3-dimensional projective spaces.

Each choice of manifold corresponds to a different physical theory.
All of the theories have, in principle, the same phenomenology near the vacuum.
However, they may present other local features at certain points, including full or partial restoration of the symmetry.
The global topology of the manifolds gives rise to a wide variety of topological defects and textures, depending on the selected manifold.
To study these effects experimentally, it is necessary to probe field values far away from the vacuum.
This is typical of cosmological scenarios, but it might also be possible at colliders if sufficiently high energies are reached.

\appendix

\begin{acknowledgments}
I would like to thank Nieves Álamo and Manuel Pérez-Victoria for helpful comments. This work has received funding from MICIU/AEI/10.13039/ 501100011033 and ERDF/EU, under the grant PID2022-139466NB-C22, and the grant RYC2021-030842-I from the Ramón y Cajal program.
\end{acknowledgments}

\section{Actions of the electroweak group}
\label{app:actions}

The classification in Section~\ref{sec:electroweak} is given in terms of actions of the group $SU(2)$.
Let us now see how to recover the action of the electroweak symmetry group $SU(2)_L \times U(1)_Y$ on each of the manifolds in the classification.
First, the principal isotropy is the diagonal subgroup $U(1)_{\rm em}$ with elements of the form
\begin{equation}
  (\operatorname{diag}(e^{i\theta}, e^{-i\theta}), e^{i\theta}) \in SU(2)_L \times U(1)_Y.
\end{equation}
The singular isotropies $K$ for each type of singular orbit are: for fixed points, $K = SU(2)_L \times U(1)_Y$; for $\mathbb{R}P^3$, $K = U(1)_{\rm em} \times \mathbb{Z}_2$, where the non-trivial element of $\mathbb{Z}_2$ acts as $(1_{2\times 2}, -1)$; and for $S^2$, $K = U(1)^2_{\rm max}$, the maximal torus, with elements of the form $(\operatorname{diag}(e^{i\alpha}, e^{-i\alpha}), e^{i\beta})$.

To construct an action of $SU(2)_L \times U(1)_Y$ on 4-dimensional manifolds, consider first its action on $S^3$ defined by the restriction of a linear representation over $\mathbb{R}^4 \cong \mathbb{C}^2$ to a 3-sphere around the origin.
For $SU(2)_L$, this linear representation is the fundamental one, while $U(1)_Y$ acts by scalar multiplication.
That is, for $g \in SU(2)_L$, $e^{i\theta} \in U(1)_Y$ and $z, w \in \mathbb{C}$,
\begin{equation}
  \begin{pmatrix}
    z \\ w
  \end{pmatrix}
  \mapsto
  e^{i\theta} g
  \begin{pmatrix}
    z \\ w
  \end{pmatrix}.
\end{equation}
When $|z|^2 + |w|^2 = 1$ both $(z, w)$ and its image live in $S^3$.

On $S^1 \times S^3$ and $\mathbb{R} \times S^3$, the group acts only on the $S^3$ factor in this way.
For the cases with $[0, 1]$ orbit space, notice that they can be covered with two open subsets, each containing one of the singular orbits and being diffeomorphic to one of the $[0, \infty)$ manifolds.
On both of these subsets, the action is given by the action on the corresponding $[0, \infty)$ manifold.

It remains to see how the group acts on the three $[0, \infty)$ cases.
On $\mathbb{R}^4$, the action is the linear representation defined in the previous paragraph.
On $\mathbb{R}P^4 \setminus \{*\}$, it can be constructed from $[0, \infty) \times S^3$ by identifying points $(0, n) \sim (0, -n)$ with $n\in S^3$.
The natural action on $[0, \infty) \times S^3$ then gives an action on $\mathbb{R}P^4 \setminus \{*\}$.
This is well defined because of the linearity of the identification
\begin{equation}
  e^{i\theta} g \cdot (-n) = -e^{i\theta} g \cdot n.
\end{equation}
The action on $\mathbb{C}P^2 \setminus \{*\}$ is obtained similarly, using the identification on $[0, \infty) \times S^3$ given by $(0, n) \sim (0, e^{i \theta} g n)$, for $(g, e^{i\theta}) \in U(1)^2_{\rm max}$.
This is also well defined because of linearity.

\section{Homotopy groups}
\label{app:homotopy}

\begin{table}
  \centering
  \begin{tabular}{ccccc}
    \toprule
    & $\pi_1$ & $\pi_2$ & $\pi_3$ & $\pi_4$ \\
    \midrule
    $S^1$ & $\mathbb{Z}$ & 0 & 0 & 0 \\
    $S^2$ & 0 & $\mathbb{Z}$ & $\mathbb{Z}$ & $\mathbb{Z}_2$ \\
    $S^3$ & 0 & 0 & $\mathbb{Z}$ & $\mathbb{Z}_2$ \\
    $S^4$ & 0 & 0 & 0 & $\mathbb{Z}$ \\
    \bottomrule
  \end{tabular}
  \caption{Homotopy groups of spheres.}
  \label{tab:homotopy-spheres}
\end{table}

The construction of the homotopy groups in Table~\ref{tab:homotopy} is based on the well-known homotopy groups of spheres~\cite{hatcher2002algebraic} listed in Table~\ref{tab:homotopy-spheres}.
The homotopy groups of projective spaces can be directly derived from them using the following relations:
\begin{align}
  \pi_n(\mathbb{R}P^m) &= \left\{\begin{array}{ll} \mathbb{Z}_2 & \text{if } n = 1 \\
    \pi_n(S^m) & \text{if } n > 1\end{array}\right.,\\
    \pi_n(\mathbb{C}P^m) &= \left\{\begin{array}{ll} \mathbb{Z} & \text{if } n = 2 \\
      \pi_n(S^{2m+1}) & \text{if } n \neq 2\end{array}\right.,
\end{align}
for $m > 1$.

For punctured real projective spaces, one can use two alternative constructions of $\mathbb{R}P^m$: identifying opposite points in an $m$-dimensional sphere $S^m$, or identifying opposite points in the boundary of a closed $m$-dimensional ball $D^m$.
The boundary of the ball is $\partial D^m = S^{m-1}$, so after the identification it becomes $S^{m-1}/\mathbb{Z}_2 \cong \mathbb{R}P^{m-1}$.
Removing one point in the interior of the $D^m$ allows to retract the resulting manifold to this $\mathbb{R}P^{m-1}$ submanifold.
Therefore,
\begin{equation}
  \pi_n(\mathbb{R}P^m \setminus \{*\}) = \pi_n(\mathbb{R}P^{m-1}).
\end{equation}

Similarly, for the punctured complex projective space, consider the identifications $\mathbb{C}P^m \cong S^{2m+1} / U(1)$ and $\mathbb{C}P^m \cong D^{2m} / U(1)$, where $U(1)$ acts by rotations on the whole $S^{2m+1}$ in the first case, and on the boundary $S^{2m-1}$ in the second one.
By removing a point in $D^{2m}$, one then gets:
\begin{equation}
  \pi_n(\mathbb{C}P^m \setminus \{*\}) = \pi_n(\mathbb{C}P^{m-1}).
\end{equation}

Finally, I have made use of the following formulas for manifolds constructed as products or connected sums of others:
\begin{align}
  \pi_n(M \times N) &= \pi_n(M) \times \pi_n(N), \\
  \pi_1(M \# N) &= \pi_1(M) * \pi_1(N),
\end{align}
where $G*H$ denotes the free product of groups, whose elements are the words of the form $g_1 h_1 g_2 h_2 \cdots g_n h_n$ for $g_i \in G$ and $h_i \in H$.
The only non-trivial free product that appears in this paper is $\mathbb{Z}_2 * \mathbb{Z}_2$, which is the infinite dihedral group $D_\infty$.
It can be viewed as a semidirect product $D_\infty = \mathbb{Z} \rtimes \mathbb{Z}_2$, with elements given by pairs $(Q, P) \in \mathbb{Z} \times \mathbb{Z}_2$ and group operation
\begin{equation}
  (Q_1, P_1) + (Q_2, P_2) = (Q_1 + (-1)^{P_1} Q_2, P_1 + P_2).
\end{equation}

\section{Lagrangian near the vacuum}
\label{app:lagrangian}

The global topology of $M$ does not impose any constraints on the Lagrangian in a neighborhood of the vacuum submanifold.
In other words, any Lagrangian compatible with the electroweak symmetry in such a neighborhood can be extended smoothly to the complete manifold $M$. One can see this through the following construction.

In $(h, U)$ coordinates, with the vacuum at $h = 0$, consider an arbitrary Lagrangian $\mathcal{L}^{\rm loc}$ of the form in Eq.~\eqref{eq:generic-lagrangian}, defined locally for $(h, U) \in (-\delta, \delta) \times S^3$.
This amounts to specifying the functions $V^{\rm loc}(h)$ and $g^{\rm loc}_{ij}(h)$ in Eq.~\eqref{eq:V-and-g}.
Additionally, let $V^{\rm glo}(\phi)$ be an invariant function, and $g^{\rm glo}(\phi)$ an invariant metric, both defined in the complete manifold $M$.
Such objects exist for all $M$ in the classification~\cite{galaz2018cohomogeneity}.
In the neighborhood $(-\delta, \delta) \times S^3$ of the vacuum, they can also be written in terms of functions of $h$ only: $V^{\rm glo}(h)$ and $g^{\rm glo}_{ij}(h)$.

\begin{figure}
    \includegraphics[width=\linewidth]{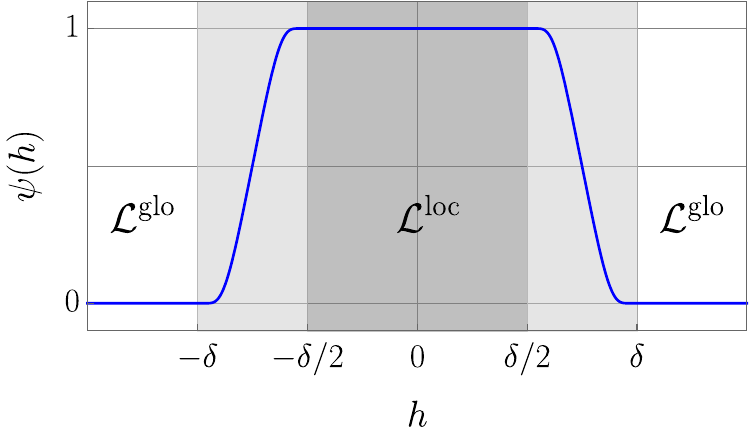}\vspace{-10pt}
    \caption{Smooth transition function $\psi$, with $\psi(h) = 0$ for $|h|>\delta$ and $\psi(h) = 1$ for $|h|<\delta/2$.
    In the darker gray region, the Lagrangian $\mathcal{L}$ constructed using Eqs.~\eqref{eq:mixed-V} and \eqref{eq:mixed-g} is identical to an arbitrary locally-defined $\mathcal{L}^{\rm loc}$. In the white regions, it is identical to the globally-defined $\mathcal{L}^{\rm glo}$. In between, it interpolates smoothly between them.}
    \label{fig:psi}
\end{figure}

Now, consider the Lagrangian $\mathcal{L}$ defined everywhere in $M$ in the following way.
Outside of $(-\delta, \delta) \times S^3$, $\mathcal{L} = \mathcal{L}^{\rm glo}$. That is, the potential and the metric are given by $V = V^{\rm glo}$ and $g = g^{\rm glo}$.
Inside of it, they are
\begin{align}
    V(h) &= \psi(h) V^{\rm loc}(h) + [1 - \psi(h)] \, V^{\rm glo}(h),
    \label{eq:mixed-V} \\
    g_{ij}(h) &= \psi(h) g^{\rm loc}_{ij}(h) + [1 - \psi(h)] \, g^{\rm glo}_{ij}(h),
    \label{eq:mixed-g}
\end{align}
where $\psi(h)$ is the smooth transition function displayed in Fig.~\ref{fig:psi} and given by
\begin{align}
    \psi(h) &= \alpha\left(\frac{\delta + h}{\delta/2}\right)
               \alpha\left(\frac{\delta - h}{\delta/2}\right), \\
    \alpha(x) &= \frac{\beta(x)}{\beta(x) + \beta(1 - x)}, \\
    \beta(x) &= \left\{\begin{array}{ll} e^{-1/x} & \text{if } x > 0\\ 0 & \text{if } x \leq 0 \end{array}\right.
\end{align}
The potential and metric defined in this way are smooth and invariant under the action of the electroweak symmetry.
Furthermore, $V(h) = V^{\rm loc}(h)$ and $g_{ij}(h) = g^{\rm loc}_{ij}(h)$ for $h \in (-\delta/2, \delta/2)$.
This proves that the choice of Lagrangian in $(-\delta/2, \delta/2)$ is independent of the topology of $M$ (which only imposes conditions on $V^{\rm glo}$ and $g^{\rm glo}$).

This procedure can be applied to the neighborhood of any principal orbit, so the functions $V(h)$ and $g_{ij}(h)$ are arbitrary (with $g_{ij}(h)$ symmetric and positive-definite) everywhere in $M$ except at singular orbits, which is where the $(h, U)$ coordinates are well-defined.

Near a singular orbit, one may use a similar set of coordinates $(H, U)$, with $H \in [0, \delta)$, $U \in SU(2) \cong S^3$ and the singular orbit at $H = 0$.
This is a generalization of polar coordinates with radius $H$.
All the pairs $(0, k U)$ with $k$ in the singular isotropy group $K$ represent the same point in $M$.
Because of this, not all smooth functions $V(H)$ and $g_{ij}(H)$ give rise to smooth functions in a neighborhood of the singular orbit.
For $V(H)$, it is necessary and sufficient that its odd derivatives vanish at $H = 0$ (as it happens in the usual polar coordinates).
Similarly, the metric is
\begin{equation}
    g = dH^2 + g_{ij}(H) dL^i dL^j,
\end{equation}
with $dL$ defined in Eq.~\eqref{eq:dL-def}, and $g_{ij}(H)$ being a positive-definite symmetric matrix with all odd derivatives vanishing at $H = 0$.
In addition to this, the elements corresponding to directions along the action of the singular isotropy must be zero at $H = 0$.
That is, for $K = SU(2)$, $g_{ij}(0) = 0$.%
The SM is a particular case where $g_{ij}(H)$ is the standard metric of the 3-sphere with radius $H$.
For $K = U(1)$, assuming it acts as $e^{i\sigma_3 \alpha}$, the condition is $g_{3i}(0) = g_{i3}(0) = 0$.

\section{UV completion example}
\label{app:example}

Consider a theory with a real scalar singlet $s$ and an $SU(2)$ doublet, viewed as real tuple $\varphi \in \mathbb{R}^4$, with Lagrangian
\begin{align}
  \mathcal{L}_{\rm UV} &= \frac{1}{2} (\partial_\mu s)^2 + \frac{1}{2} |\partial_\mu \varphi|^2
  \nonumber \\
  &\phantom{=} - W(s) - \omega (s^2 + |\varphi|^2 - u^2)^2,
\end{align}
where $\omega$ and $u$ are free parameters and $W(s)$ is some polynomial potential for $s$.
The pair $(s, \varphi)$ is an element of $\mathbb{R}^5$.
Choosing a spherical parametrization, one has $(s, \varphi) = \rho n$, with $n \in S^4$ a unit 5-vector.
In these coordinates, the Lagrangian is
\begin{align}
  \mathcal{L}_{\rm UV} &= \frac{1}{2} (\partial_\mu \rho)^2 + \frac{\rho^2}{2} g_{ij}(n) \partial_\mu n_i \partial^\mu n_j \nonumber \\
                       &\phantom{=} - W(\rho n_1) - \omega (\rho^2 - u^2)^2,
\end{align}
where $g$ is the standard metric of the unit $4$-sphere.
When $W(s)$ is constant, the theory has an $SO(5)$ global symmetry, which is broken down spontaneously to $SO(4)$.
So, the EFT for the Goldstones will have a scalar target space $SO(5) / SO(4) \cong S^4$.

Let us now see what happens when $W(s)$ is not constant.
In the limit $\omega \to \infty$ with all other parameters fixed, the radial field is frozen to $\rho = u$.
The resulting effective Lagrangian is
\begin{align}
  \mathcal{L} &= \frac{u^2}{2} g_{ij}(n) \partial_\mu n_i \partial^\mu n_j  - W(u n_1).
\end{align}
This is equivalent to integrating out the radial perturbation $\delta\rho = \rho - u$ at tree level and taking its mass $M^2 = 8 \omega u^2$ to infinity.
If $\omega$ was kept finite, the effective Lagrangian would receive corrections proportional to inverse powers of $M$.

The scalar target space is again $S^4$.
Both the metric and the potential are smooth and invariant under the action of $SU(2)$.
All orbits are 3-spheres, except for the two fixed points, located at $n_1 = \pm 1$.
The potential determines the vacuum submanifold.
For concreteness, I will choose here the following form for it
\begin{equation}
  W(s) = \mu^2 \left(\frac{s^2}{2} - u s \cos\theta\right),
\end{equation}
with two free parameters $\mu$ and $\theta$.
The vacuum is then the 3-sphere given by the condition $n_1 = \cos\theta$.
One can change to the usual $(h, U)$ variables as
\begin{align}
  h &= \frac{\sqrt{1 - n_1^2}}{\cos\theta} - \tan\theta, \\
  U &= \frac{1}{\sqrt{1 - n_1^2}}
  \begin{pmatrix} n_2 + i n_3 & n_4 + in_5 \\ -n_4 + i n_5 & n_2 - i n_3 \end{pmatrix}.
\end{align}
The relation between these and the original variables is smooth and invertible in some neighborhood of the vacuum, which is now at $h = 0$.
Ignoring constant terms, the Lagrangian becomes
\begin{align}
  \mathcal{L} &= \frac{1}{2} F_h(h)^2 \partial_\mu h \partial^\mu h - V(h) \nonumber \\
  &\phantom{=} + \frac{v^2}{4} F_U(h)^2 \langle\partial_\mu U \partial^\mu U^\dagger\rangle,
\end{align}
where $v = u \sin\theta$, and
\begin{align}
  F_h(h)^2 &= \frac{\cos^2\theta}{1 - (\sin\theta)^2 (1 + (h/v)\cos\theta)^2}, \\
  F_U(h) &= 1 + \frac{h}{v} \cos\theta, \\
  V(h) &= \frac{\mu^2}{2} (\sin\theta)^2 h^2 + O(h^3).
\end{align}

Taking $v = 246 \, \text{GeV}$, the experimental measurements of the Higgs mass and its couplings to gauge bosons~\cite{ParticleDataGroup:2024cfk} require%
\footnote{Ref.~\cite{ParticleDataGroup:2024cfk} obtains, from a weighted average of the available data, and interval $1.023\pm 0.026$ for the Higgs coupling to gauge bosons, which implies the limit Eq.~\eqref{eq:cos-limit}.}
\begin{align}
  \cos \theta &> 0.997, \label{eq:cos-limit}\\
  \mu \sin \theta &= 125 \, \text{GeV}.
\end{align}
These conditions are satisfied, for example, for $\theta = 0.05$, $\mu \simeq 2.5\,\text{TeV}$, and $u \simeq 5\,\text{TeV}$.

Finally, I will illustrate the global effects described in Section~\ref{sec:global} in the context of this example.
According to Table~\ref{tab:homotopy}, the only ones that arise here are the spacetime textures, associated with the fourth homotopy group $\pi_4(S^4) \cong \mathbb{Z}$.
Since the $(h, U)$ coordinates are only valid in a neighborhood of the vacuum, I return here to the $n$ parametrization.
In order for a spacetime configuration $n(x)$ of the fields to have finite action, it is necessary that it goes to a fixed point $n_{\infty}$ in the vacuum submanifold at spacetime infinity.
The topological charge associated with $n(x)$ is
\begin{equation}
    Q = \frac{\epsilon^{\mu\nu\rho\sigma} \epsilon^{ijklm}}{8\pi^2 / 3} \int d^4x \,
    n_i \partial_\mu n_j \partial_\nu n_k \partial_\rho n_l \partial_\sigma n_m.
\end{equation}
The integral on the right-hand side measures the signed volume swept by $n(x)$ on the unit 4-sphere, with positive sign if $n(x)$ preserves the orientation and negative sign otherwise.
Because, in this setup, spacetime can be compactified to $S^4$, the swept volume can only be an integer multiple of the volume $8\pi^2/3$ of the unit 4-sphere.
This ensures that $Q$ is an integer.

An example of a configuration with $Q=1$ is given by the inverse of the stereographic projection from $n_\infty = (c_\theta, s_\theta, 0, 0, 0)$:
\begin{equation}
    n(x) = \frac{1}{|x|^2 + 1}
    \left(
    \begin{array}{c}
    (|x|^2 - 1) \cos\theta - 2 x_0 \sin\theta \\
    (|x|^2 - 1) \sin\theta + 2 x_0 \cos\theta \\
    2 x_1 \\
    2 x_2 \\
    2 x_3
    \end{array}
    \right)
\end{equation}
where $|x|^2 = \sum_i x_i^2$ is the Euclidean distance to the origin.
This configuration represents a smooth topologically non-trivial process starting with a homogeneous spatial configuration $n(\mathbf{x}) = n_\infty$ at the infinite past and returning to the same one in the infinite future.

At the classical level, such a process cannot take place, because it goes through a barrier of spatial configurations $n(\mathbf{x})$ with higher energy than the vacuum.
At the quantum level, the tunneling rate could in principle be computed semiclassically by extremizing the Euclidean action over smooth deformations of the inverse stereographic projection.
Determining if this is possible and deriving the associated phenomenology will require further work.

\vfill

\bibliography{references}

\end{document}